\documentclass[prb,twocolumn,showpacs,preprintnumbers,amsmath,amssymb]{revtex4}
\usepackage{graphicx}% Include figure files
\usepackage{dcolumn}% Align table columns on decimal point
\usepackage{bm}% bold math

\usepackage{color}

\newcommand{\grad}{{ \nabla }}

\newcommand{\infinity}{\infty}

\newcommand{\ignore}[1]{}  % a command which does nothing.
\newcommand{\add}[1]{#1}

\newcommand{\del}[1]{}

\newcommand{\eqr}[1]{Eq.~(\ref{#1})}

\newcommand{\vB}{\bm{B}}

\newcommand{\vR}{\bm{R}}
\newcommand{\vS}{\bm{S}}

\newcommand{\vZ}{\bm{Z}}

\newcommand{\vj}{\bm{j}}

\newcommand{\vu}{\bm{u}}

\newcommand{\vx}{\bm{x}}

\newcommand{\vecphi}{\bm{\phi}}

\newcommand{\vell}{{\vec \ell}}
\begin{document}

\preprint{PPPL-3788 arXiv:Physics/0303026}

%\title{On the existence of tokamak equilibria with negative central current}
\title{Non-existence of normal tokamak equilibria with negative central
current}

\author{G. W. Hammett\footnote{Electronic mail: hammett@princeton.edu}}
\author{S. C. Jardin}
\author{B. C. Stratton}
%\email{hammett@princeton.edu}
\affiliation{Princeton Plasma Physics Laboratory}
%\homepage{http://w3.pppl.gov/~hammett}

\date{Feb. 14, 2003, Rev. May 28, 2003, accepted for publ. in Phys. Plasmas}
%\date{\today , submitted to Phys. Plasmas }% It is always \today, today,
% don't use \today for papers submitted to www.arxiv.org.

\begin{abstract}
Recent tokamak experiments employing off-axis, non-inductive current
drive have found that a large central current hole can be produced.  The
current density is measured to be approximately zero in this region,
though in principle there was sufficient current drive power for the
central current density to have gone significantly negative.  Recent
papers have used a large aspect-ratio expansion to show that normal
Magnetohydrodynamic (MHD)
equilibria (with axisymmetric nested flux surfaces, non-singular fields,
and monotonic peaked pressure profiles) can not exist with negative
central current.  We extend that proof here to arbitrary aspect ratio,
using a variant of the virial theorem to derive a relatively simple
integral constraint on the equilibrium.  However, this constraint does
not, by itself, exclude equilibria with non-nested flux surfaces, or
equilibria with singular fields and/or hollow pressure profiles that may
be spontaneously generated.
%
% Tokamak plasmas with low central current density, and hence reversed
% magnetic shear in the center, are of interest because they have
% been observed with internal transport barriers and because they are
% compatible with high-bootstrap current and low current drive requirements.
\end{abstract}

\maketitle

\section{\label{sec:Introduction}Introduction}

% Virial theorem due to Clausius (1800's).  States that 
%
% -<r.F>/2 = K.E.
%
% where r is the position vector, F is the force, and <...> is an
% average of some sort (time average of many particles or, in our case,
% spatial average).

%%%%%%%%%%%%%%%%%%%
%
% Stratton modification:
%

Tokamaks with reversed central magnetic shear (and thus low core current
density) are of interest for at least two reasons: 1) internal transport
barriers 
associated with reduced turbulence are often observed in them,
leading to improved 
% comment out if needed to save space:
energy and particle 
confinement; and 2) they are the natural result of high beta operation
and high bootstrap current fraction used to reduce non-inductive
current drive requirements for steady state operation.  Both of these
features could make reversed magnetic shear operation attractive for a
tokamak reactor.

Recent experiments on the Joint European Torus
(JET)\cite{Hawkes2001,Stratton2002,Hawkes2002,Stratton2002b} and the
Japan Atomic Energy Research Institute Tokamak-60 Upgrade
(JT-60U)\cite{Fujita2001,Miura2002,Chankin03} have pushed the core
current density to very low values using off-axis, non-inductive current
drive. Large central current holes (regions of nearly zero current
density) are produced because off-axis, non-inductive current drive in
the same direction as the Ohmic current induces a back electromotive
force inside the non-inductive current drive radius that decreases the
core current density.
%
% Not directly important to this paper, so save space:
%
% The non-inductive current drive used in the JET
% experiments is lower hybrid current drive, and bootstrap current
% generated by neutral beam injection is used in the JT-60U experiments.

An interesting feature of current hole discharges is that the core
current density is approximately zero (within Motional Stark Effect
diagnostic measurement errors), even though there is often sufficient
current drive power that the core current could in principal go
significantly negative\cite{Stratton2002,Hawkes2002} (negative relative
to the direction of the total plasma current). Recent non-linear
toroidal resistive MHD (Magnetohydrodynamic)
simulations\cite{Stratton2002,Breslau2002}
predict that current hole discharges undergo rapid $n=0$ reconnection
events (axisymmetric sawteeth) that clamp the core current near zero.
More generally, this reconnection occurs whenever the current density
profile is such that the rotational transform, $\iota$, goes to zero on
any surface in the plasma (this includes the case where the current
density on-axis is positive, but the current profile goes sufficiently
negative somewhere off-axis that the total current enclosed by some flux
surface vanishes).  
Reduced MHD simulations in cylindrical geometry have also shown that
$n=0$ resistive kink instabilities can clamp the core current density at
zero when it attempts to go negative.\cite{Huysmans2001}
Breslau {\it et al}.\cite{Breslau2002}\ and Stratton {\it et
al}.\cite{Stratton2002}\ stated that a second-order, large
aspect ratio expansion of the MHD equations indicates that a normal
toroidal equilibrium is not possible if $\iota$ crosses through $0$ at some
radius. (They also stated that a more general proof is needed, which we
provide here.)  A recent paper by Chu and Parks\cite{Chu2002} used
a second order aspect ratio expansion to prove that a normal equilibrium
with a peaked pressure profile is not possible with negative core
current. They extended the analysis to provide matching conditions at
the boundary of a central region with no current and no pressure
gradient, showing explicitly that current hole equilibria are
theoretically possible (with zero, but not negative, current). 
% They also
% demonstrated that equilibria with non-monotonic pressure profiles are
% isolated solutions of the force balance equation and therefore can not
% be experimentally realized.

This paper extends some of these results to arbitrary aspect ratio,
employing a relatively simple constraint based on a version of the
virial theorem to show that a ``normal'' toroidal MHD equilibrium (with
axisymmetric nested flux surfaces around a single magnetic axis,
non-singular continuous fields, and a monotonic peaked pressure profile)
is not possible with negative core 
current.  Or more generally, a normal equilibrium is not possible if the
toroidal current enclosed by any flux surface goes negative relative to
the direction of the total plasma current, so that there is an $\iota=0$
surface somewhere in the plasma where the poloidal field vanishes (the
null surface). 
Though the starting point of this analysis is based on well-known
equations, they are often specialized to large aspect ratio or
simplified geometry, while the present analysis is more general. 
% Amount of space needed to cut to get to 4 pages:
% \ignore{

However, the virial constraint does not necessarily eliminate the
possibility of more exotic equilibria, such as with non-nested flux
surfaces with islands, or with singular fields and/or off-axis peaks in
the pressure profiles that may 
be spontaneously generated by the plasma near the null surface.  Some
examples are considered here.  In this paper we investigate the
consequences of only one integral constraint on equilibria, while there
can be other constraints that further limit the types of theoretically
possible or experimentally realizable equilibria\cite{Chu2002}.

% We also speculate on the possibility of an equilibrium with non-nested but
% still axisymmetric flux surfaces, and give an example with two islands.

% }

% Sometimes put section referring to Westerhof here to fit on 4 pages:
% Would prefer to put this in the summary, but the spacing gets messed
% up and runs to a 5th page:
The non-existence of normal equilibria with negative core current,
and/or the rapid axisymmetric sawteeth that are predicted to occur if
the enclosed current goes negative, may also explain the results of other
experiments, such as the low efficiency seen in some electron cyclotron
counter current drive experiments\cite{Westerhof2001,Miura2002}.

%
% END Stratton modification:
%
%%%%%%%%%%%%%%%%%%%

% Friedberg, and even Shafranov, do large aspect ratio expansions.

%%%%%%%%%%%%
%

% The virial theorem as used in Freidberg dots the MHD force balance
% equation with $\vec r$ and integrates over space to show that isolated
% equilibrium without external coils is impossible.  Another version of
% the virial theorem dots the MHD force balance equation with $\vec R$
% (the major radius vector in cylindrical coordinates) and integrates over
% the plasma volume inside a specified flux surface.  Rutherford used this
% in his tokamak class when I took it to derive the Shafranov shift.  He
% assumed circular flux surfaces for the latter part of his derivation,
% but one can generalize this for arbitrary flux surface shape to get this
% constraint for current hole plasmas.

%%%%%%%%%%%%%%%%%%%%%%%%%%%%%%%%%%%%%%%%%
\section{\label{sec:Derivation}Derivation}

\begin{figure}
\includegraphics[width=2.8in]{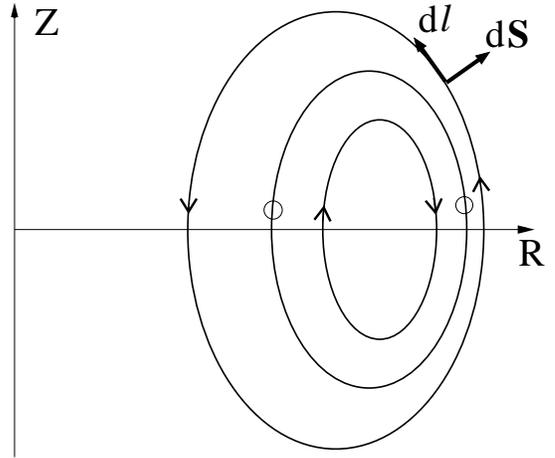}%
\caption{Sketch of hypothetical equilibrium with nested flux surfaces
and negative central current (relative to the total current), so that
the poloidal field points clockwise near the axis, counterclockwise near
the edge, and is zero on a flux surface in between.  A normal
equilibrium is not possible in this case (with a normal peaked pressure
profile)\label{fig:fig1}}
\end{figure}

The MHD equilibrium equation $\grad p = {\vj} \times \vB / c$ can
be written as
\begin{equation}
0 = - \grad \left( p + \frac{B^2}{8 \pi} \right) + \frac{1}{4 \pi}
\left( \vB \cdot \grad \right) \vB
\label{MHD-eq}
\end{equation}
One common use of the virial theorem is to take the inner product of
this equation with the
position vector $\vx$ and integrate over all space to show that an
isolated MHD equilibrium can not exist by itself (unless there are
physical coils or gravity to provide overall force
balance).\cite{Shafranov66, Freidberg87} Here we use a version of the
virial theorem that can be used to derive the
Shafranov shift\cite{Rutherford81}, by focusing on radial force balance
of axisymmetric equilibria in cylindrical coordinates $(R,Z,\phi)$.
Taking the inner product of \eqr{MHD-eq} with
${\vR} = R {\hat \vR}$, the radial
vector in cylindrical geometry, and integrating over space out to some
flux surface of volume $V$, gives
\begin{equation}
0 = - \int d V {\vR} \cdot \grad \left( p + \frac{B^2}{8 \pi} \right)
+ \frac{1}{4 \pi} \int d V {\vR} \cdot \left( \vB \cdot \grad
\right) \vB
\label{eq:rdotMHD}
\end{equation}
For the second integral we use the identity $\vR \cdot (\vB \cdot
\grad ) \vB = \grad \cdot (\vB \vR \cdot \vB) - B_R^2 - B_\phi^2$.  The
integral of $\grad \cdot (\vB \vR \cdot \vB)$ vanishes because $\vB
\cdot d {\vS} = 0 $ on a flux surface.  The first integral can be
integrated by parts using $\grad \cdot \vR = 2$, so that
\eqr{eq:rdotMHD} becomes
\begin{eqnarray}
0 & = &-p(\rho) \int d \vS \cdot \vR 
      - \int d \vS \cdot \vR \frac{B^2}{8 \pi} \nonumber \\
& &  + 2 \int d V \left( p + \frac{B^2}{8 \pi} \right)
    - \frac{1}{4 \pi} \int d V \left( B_R^2 + B_\phi^2 \right) \nonumber \\
& = & -p \int \! d \vS \cdot \vR
      - \int d \vS \cdot \vR \frac{B^2}{8 \pi} 
  + 2 \int d V \left( p + \frac{B_Z^2}{8 \pi} \right) , \nonumber \\
\end{eqnarray}
%
% Could have simplified the first surface integral already in the above
% equation, but I wanted to leave Eq.3 close to the form that Rutherford
% had in his course notes, to help differentiate my modest
% contributions.  From this point out, Rutherford's notes focus on
% cylindrical large-aspect ratio geometry.
%
where $p=p(\rho)$ is the pressure at the surface labeled by $\rho$
enclosing the volume $V(\rho)$, and $B_Z$ is the vertical magnetic field.
For the first surface integral we can use Gauss' theorem to 
write $\int d \vS \cdot \vR = \int dV \grad \cdot \vR = 2 V$. For the
second surface integral, we use $d \vS = 2 \pi R \hat{\vecphi}
\times d \vell $, where $d \vell$ is a poloidal path length element
along the surface,  to write $d \vS \cdot \vR = 2 \pi R^2 {\hat{\vZ}}
\cdot d \vell $ so $\int d \vS \cdot \vR = 2 \pi \oint R^2 {\hat{\vZ}}
 \cdot d \vell = 2 \pi \oint R^2 dZ$. Since the toroidal field
$B_\phi \propto 1/R$ in a flux surface, the $B_\phi^2$ contribution to
this surface integral vanishes, and we have 
\begin{equation}
0 = - p(\rho) - \frac{1}{V} \oint d \vell \cdot \hat{\vZ} \pi R^2
\frac{ {B_{pol}^2} }{8 \pi}
+ \langle p \rangle + \left\langle \frac{ B_Z^2 }{8 \pi} \right\rangle
\label{eq:virial-gen}
\end{equation}
where $\langle \ldots \rangle = \int dV \ldots / \int dV$ denotes a
volume average, and $B_{pol}^2 = B_R^2 + B_Z^2$ is the poloidal field
strength squared.  The poloidal field can be written as $\vB_{pol} =
\grad \phi \times \grad \psi = (\hat{\vecphi} \times \grad \rho)
(\partial \psi / \partial \rho) /R$, where $\rho$ is a flux surface
label.  [While $\psi$ is also constant on a flux surface, there can be
two surfaces with the same value of $\psi$ in the presence of negative
central current, so it \add{is} 
convenient to choose another flux surface label
$\rho$, such as based on the enclosed volume or toroidal flux, to
maintain monotonic labeling.]  If the toroidal current near the magnetic
axis is in the opposite direction as the total plasma current, then the
poloidal field must reverse direction somewhere and there must be a null
surface on which the poloidal magnetic field is everywhere zero, as
shown in Fig.~\ref{fig:fig1}.  Another way to see this is to note that
the poloidal field is related to the enclosed toroidal current by $4 \pi
I_\phi(\rho)/c = \oint d \vell \cdot \vB_{pol} = (\partial \psi /
\partial \rho) \oint d \vell \cdot \hat{\vecphi} \times \grad \rho / R$,
so $B_{pol} \propto \partial \psi / \partial \rho = 0$ on any flux
surface that encloses zero toroidal current.  This is also the flux
surface on which the rotational transform $\iota = 0$ (corresponding to
the safety factor $q=\infinity$).  [These arguments assume that
$B_{pol}$ is continuous and finite, we consider a singular exception in
the next section.]
%
% Josh Breslau suggests the following for getting iota-bar.
% It works, but more work is needed to get the spacing for it right.
%
% \begin{picture}(1,4)(0,0)
% \put (0,0){$\iota = 0$,}
% \put (0,0){-}
% \end{picture}
%
% Jardin says many people just define iota = 1/q now, so that is what we
% will do.

On such a flux surface where the poloidal field vanishes, the second
term of \eqr{eq:virial-gen} vanishes and we are left with constraint
\begin{equation}
0 = -p + \left\langle p \right\rangle + \left\langle \frac{B_Z^2}{8 \pi}
\right\rangle
\label{eq:virial-null}
\end{equation}
Since the last two terms are positive definite, the only way this
equation can possibly be satisfied is if the pressure at this flux
surface, $p$, is larger than the volume-averaged pressure inside that
flux surface, $\langle p \rangle$.  I.e., the pressure profile must be
hollow \add{at least in some region}, and \add{can not be a
monotonically decreasing function of $\rho$ at all radii}
\del{can't monotonically decrease with increasing $\rho$}%
as usual pressure profiles do.  

This is in agreement with the result of Chu and Parks\cite{Chu2002}, who
also found that a
normal equilibrium with a negative central current is not possible if
the pressure profile is monotonically decreasing.  These earlier results
used a second-order large aspect ratio ordering while our derivation is
valid for arbitrary aspect ratio.  In other ways, their calculation goes
beyond ours, as they have investigated additional constraints that can
further limit the class of accessible equilibria.

% end ignore

\begin{figure}
\includegraphics[width=2.8in]{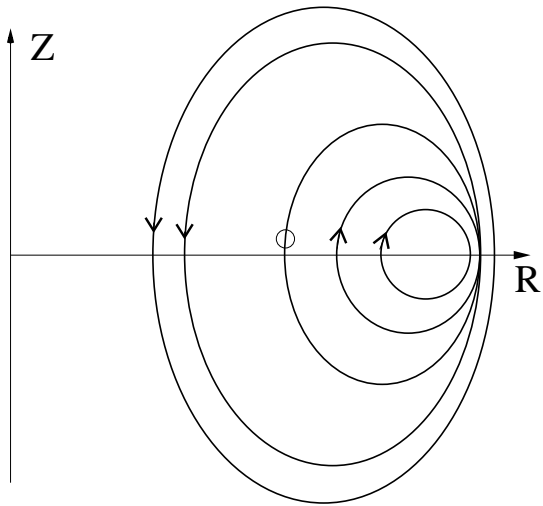}%
\caption{Ideal MHD \add{formally} allows singular solutions where the
poloidal field vanishes almost everywhere on a flux surface, so that the
enclosed current $\propto \oint d \vell \cdot \vB_{pol}=0$ but $\oint d
\ell B_{pol}^2$ is finite.  Such an equilibrium could then in principle
satisfy the integral force balance \eqr{eq:virial-gen}.  In this case,
adjacent flux surfaces approach one another 
at one point where the poloidal field
becomes infinite, but this is an integrable singularity with finite
energy. [All of these sketches are intended only to illustrate topology
and are not precise.]\label{fig:sing}}
\end{figure}

\section{Possible alternate solutions}
% Possibility of equilibria with non-nested surfaces}

Here we consider several possible alternate solutions for satisfying
force balance in equilibria.  Each differs from ``normal'' equilibria in
a different way.

\begin{figure}
\includegraphics[width=2.8in]{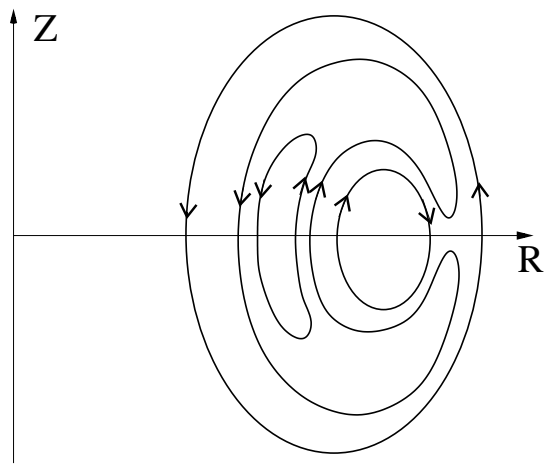}%
\caption{An equilibrium with this topology of non-nested flux surfaces 
is not ruled out by the integral force balance \eqr{eq:virial-gen}.
Note that the toroidal current in the inner part of the plasma is in the
reverse direction from the total plasma current.\label{fig:island}}
\end{figure}

The argument in the previous section
% that since the poloidal vanishes
% on a null surface enclosing no net toroidal current, so the second term
% of \eqr{eq:virial-gen} vanishes, 
applies rigorously only for nested flux-surfaces where the fields are
continuous and finite.  Within the framework of ideal MHD, in principle
there could be a singular poloidal field on a flux surface such that
$\oint d \vell \cdot \vB_{pol} = 0$ so that this surface \del{includes}%
\add{encloses} zero
toroidal current, but $\int d \vell \cdot \hat{\vZ} \pi R^2 B_{pol}^2 /
(8 \pi)$ is still finite and can contribute to \eqr{eq:virial-gen}
so that integrated force balance can be satisfied.  An example of such a
field might be the limiting case $B_{pol} \propto |\grad{\psi}| \sim C
\exp(-\ell^2/w^2)/\sqrt{w}$.  Then in the limit as $w \rightarrow 0$ we
have zero toroidal current enclosed while still giving a finite
contribution to the second term of \eqr{eq:virial-gen} \add{(due to a
singularity in the field chosen to be at $\ell=0$ in this example)}.  Although
$B_{pol}$ is becoming infinite at some point on the flux surface, it is
an integrable singularity containing a finite amount of energy, and so
could formally be considered as an admissible solution of ideal MHD.
Since the spacing between two nearby flux surfaces labeled by poloidal
flux $\psi_1$ and $\psi_2$ is given by $\Delta \approx (\psi_2 -
\psi_1)/|\grad \psi|$ (except where second derivatives have to be
considered), there will be flux surfaces with finite separation at some
places which will approach one another
at the singular point where $|\grad \psi| \add{\ \propto B_{pol}}
\rightarrow \infinity$.  The topology of this configuration is
illustrated in Fig.~\ref{fig:sing}.  [This sketch is intended only to
illustrate the topology of a possible solution which satisfies the
integral force balance constraint, \eqr{eq:virial-gen}.  An actual
detailed solution that would satisfy force balance locally at all points
is left for future work\add{, and may involve highly distorted
flux-surface shapes with boundary layers\cite{Park80,Rosenbluth73,
Cowley91}}.]  Note that not
only is the poloidal field 
infinite at the singular point, it must also flip signs from
$+\infinity$ to $-\infinity$ in the limit as the singular point is
approached radially from opposite directions.  Of course this \add{singular}
configuration is strongly susceptible to magnetic reconnection when
resistivity is included, consistent with the interpretation of rapid
reconnection observed in simulations\cite{Stratton2002,Breslau2002}.  If
finite resistivity is included, then the singularity in the field is
smoothed out and loss of equilibrium can help drive reconnecting flows.

Another possible way of satisfying integrated force balance with
negative central current, while keeping the field finite and 
continuous, is if the flux surfaces are non-nested.  An example is shown
in Fig.~\ref{fig:island}, which is similar to the intermediate
configurations observed during some toroidal simulations of axisymmetric
reconnection in negative central current plasmas (for example, Fig.~11
of Ref.~\onlinecite{Stratton2002}, though in other cases they see
islands with higher poloidal mode numbers).
%
% more traditional accounting would call this 1 island, not 2:
%
% This case has two axisymmetric islands and the
This case has \del{two}\add{an} axisymmetric island\del{s}%
and the
poloidal field is \del{finite}\add{non-zero} almost everywhere (except
at the two magnetic
axes and at the X-point) and so in principle can be arranged to give a
negative contribution to the second term of
\eqr{eq:virial-gen} to satisfy integrated force balance.  This is
related to the role in a normal equilibrium of the Shafranov shift,
which provides a larger value of $R^2 B_{pol}^2$ on the outer part of a
flux surface than on the inner part, so that the second term in
\eqr{eq:virial-gen} is negative.  
%
% Note that the toroidal current
% density in the inner island in Fig.~\ref{fig:island} is in the opposite
% direction as the toroidal current in the outer island and opposite to
% the total plasma current enclosed by the outer flux surfaces.
%
The X-point of a non-nested equilibrium might not be on the
low-field side, and another possible equilibrium might be obtained by
rotating Fig.~\ref{fig:island} by 180$^\circ$ and shifting the spacing
between flux surfaces so that the integral of $R^2 B_{pol}^2$ on the
outer part of the flux surface is again larger than on the inner part.
One or the other of these configurations may be an unstable equilibrium
and prefer to flip to the other orientation.
[Takizuka\cite{Takizuka2002} earlier proposed another possible
non-nested equilibria with negative central current, involving
$(m=2,n=0)$ islands, while the example we discussed here has an
$(m=1,n=0)$ island.]

% A straightforward observation not crucial to our argument, so save
% space:
%
% Particle banana widths, and hence neoclassical transport, formally
% becomes infinite right at the surfaces that connect to the Y-point if
% only the local poloidal field on those surfaces is used.  One would need
% to include the radial variation of the poloidal field (such as is done
% to get ``potato'' orbits near the magnetic axis of regular tokamaks) for
% a more accurate neoclassical theory in this region.

% More work is needed to investigate whether such equilibria are actually
% theoretically possible or experimentally realizable.  While
% \eqr{eq:virial-gen} does not rule out such equilibria, there may be
% other constraints that might prohibit them\cite{Chu2002}.  Or they may
% be unstable in ideal or resistive
% MHD\cite{Breslau2002,Stratton2002,Huysmans2001}.  It would be
% interesting to study these issues further.  

Another way of thinking about ideal MHD equilibria is to modify the
time-dependent ideal MHD equations to include viscosity and parallel
thermal conduction while retaining the ideal Ohm's
law\cite{Freidberg87}.  Since viscosity should eventually damp the
velocity $\vu$ to zero, and parallel thermal conduction will lead to
$\vB \cdot \grad p = 0$, the dissipative terms vanish in a stationary
steady state and the solution is also an ideal MHD equilibrium.  One can
then start with any arbitrary initial configuration of the magnetic
field (which can be assumed to be nested flux surfaces) with arbitrary
initial profiles, as functions of toroidal flux $\Phi$, for the rotational
transform $\iota(\Phi)$ and the
adiabatic parameter $\mu(\Phi)=p/n^\Gamma$ (where $p(\Phi)$ is the
pressure profile, $n(\Phi)$ is the density profile, and $\Gamma$ is the
ratio of specific heats).  Since this initial condition is not
necessarily an equilibrium, flows will be driven and the plasma will
move about, perhaps oscillating for a while.  But it seems reasonable to
assume that the viscosity will eventually damp out the oscillations
and the plasma will settle into an equilibrium configuration while
conserving $\iota(\Phi)$ and $\mu(\Phi)$.  (The motions are assumed to
be constrained to be axisymmetric to find such an equilibria.  This
approach to equilibria of course does not address the issue of
stability, and these equilibria might then be unstable to
symmetry-breaking perturbations.)  

This was the logic that motivated the
flux-conserving tokamak equilibria concepts\cite{Freidberg87} that
showed that there is formally no equilibrium limit on the pressure in a
tokamak, since as the pressure increases, the Shafranov shift and the
current can also increase to provide sufficient poloidal magnetic field
on the outboard side to provide force balance.  

Presumably this
procedure would also find an equilibrium even if the rotational
transform changed sign so that there was a null flux surface where
$\iota=0$.  
%
% Jardin's point:
%
In some cases with certain initial conditions, it might be possible for
the plasma to spontaneously 
adjust flux surfaces near the null $\iota$ surface to produce a
local peak in the pressure profile that can satisfy
\eqr{eq:virial-null}.
[We have focussed on the consequences of only one integral
constraint that rules out ``normal'' equilibria with negative central
current, and there can be other constraints that would further limit the
practical accessibility of such \del{hollow}\add{non-monotonic} pressure
equilibria\cite{Chu2002}.]
The more typical case is probably that
the equilibrium that is approached will have a singular structure,
such as in Fig.~\ref{fig:sing}, in order to satisfy
\eqr{eq:virial-gen}.
[This is similar to studies showing that the nonlinear saturation of
an internal kink mode approaches a neighbouring equilibrium state with
singular currents\cite{Park80,Rosenbluth73}.]
These singular or near-singular states will be subject to strong
reconnection if a small amount of resistivity is introduced, and the
change in topology may dominate what happens.
Realistic evaluations of what happens may depend on fully including
various dissipation mechanisms (thermal, momentum, and particle
anisotropic transport driven by small scale turbulence and collisional
effects, as well as resistivity, current drive, heating and loss
mechanisms).  We leave detailed investigation of these issues to other
work.

%%%%%%%%%%

\vspace{\parsep}  % For some reason have to force some space here!

% \section{More unusual equilibria}
% 
% \section{1-D Slab analogy}

\section{Relation to the Shafranov shift}

For completeness, we show the relation of \eqr{eq:virial-gen} to usual
expressions for the Shafranov shift\cite{Shafranov66,Freidberg87}.  The
second term of
\eqr{eq:virial-gen} can be written as
\begin{equation}
T_2 = - \frac{\pi}{V} \oint d \vell \cdot \hat{\vZ} R^2
\frac{ {B_{pol}^2} }{8 \pi}
= - \frac{\pi}{V} \frac{ (\partial \psi / \partial \rho)^2 }{ 8 \pi}
\oint d \vell \cdot \hat{\vZ} |\grad \rho|^2
\end{equation}
At this point, many previous calculations specialize to a
large aspect ratio expansion and/or to specified shapes for the flux
surfaces.  For example, assume shifted circular flux surfaces with
$R(\rho,\theta) = R_0 - \Delta(\rho) + \rho \cos \theta$ and
$Z(\rho,\theta) = \rho \sin \theta$, where $\rho$ has now been chosen to
be the minor radius of the flux surface, and $\Delta$ is the Shafranov
shift.  It can be shown that $|\grad \rho|^2 = 1 / (1 - \Delta' \cos
\theta)^2$, where $\Delta' = d\Delta/d\rho$.  Defining $\partial \psi /
\partial \rho = B_{p}(\rho) (R_0 - \Delta)$, we have $B_{pol} = B_{p}
[R_0 - \Delta]/[R(1-\Delta' \cos \theta)]$ (this would be exact if the
flux surfaces really were shifted circles), and $T_2$ becomes
\begin{equation}
T_2 = - \frac{ {B_{p}^2} }{8 \pi}
        \frac{ R_0 - \Delta}{r}
        \frac{ \Delta' }{(1-\Delta')^{3/2}}
\end{equation}
where we have used $\int d \theta \cos \theta / (1-\Delta' \cos \theta)^2
= 2 \pi \Delta' / (1-\Delta')^{3/2}$.  Inserting this into
\eqr{eq:virial-gen} yields
\begin{equation}
\frac{ \Delta' }{(1-\Delta')^{3/2}} =
\frac{r}{R_0 - \Delta} \frac{8 \pi}{B_{p}^2} 
\left( \langle p \rangle - p + \left\langle \frac{B_Z^2}{8 \pi}
\right\rangle \right)
\label{nonlinear-Shaf}
\end{equation}
Evaluating this in the large aspect ratio limit at the plasma edge
where $p=0$ gives the familiar form $\Delta' = (r/R_0) (\beta_{pol} +
\ell_i/2)$, where $\beta_{pol}$ is the poloidal beta and $\ell_i$ is the
internal inductance per unit length.
\add{[Note that $\ell_i/2 = \langle B_Z^2 \rangle /B_{p}^2(a) 
= \langle B_{pol}^2 cos^2(\theta) \rangle / B_{p}^2(a)
= \langle B_{pol}^2 \rangle / 2 B_{p}^2(a)$, so that
the factor of $1/2$ comes from the fact that only the 
vertical magnetic field contributes to the numerator in the shift.]}
The nonlinear form of the
left-hand side of \eqr{nonlinear-Shaf} has the nice property of insuring
that the flux surfaces are well behaved and don't cross ($|\Delta'| <
1$) for arbitrarily high $\beta_{pol}$, though this equation only
rigorously applies if the flux surfaces remained shifted circles, which
breaks down at high beta.

\section{Summary}

We have presented a relatively simple integral constraint on toroidally
axisymmetric MHD equilibrium that shows that a normal equilibrium (with
nested magnetic flux surfaces, non-singular fields,
and a peaked pressure profile that falls
monotonically with radius) can not exist if the toroidal current
inside any flux surface is negative relative to the total plasma
current.  This generalizes previous
results\cite{Breslau2002,Stratton2002,Chu2002} to arbitrary aspect
ratio.  

However, the integral constraint does not necessarily prevent negative
central current equilibria with non-nested or singular magnetic flux
surfaces.  Possible examples of this are shown in Figs.~\ref{fig:sing}
and \ref{fig:island}.  
A plasma with nested flux surfaces and negative central current that is
initially out of equilibrium could presumably move towards an
equilibrium, though it seems most likely that this new equilibrium would
have singular or near-singular fields and thus would be subject to
strong reconnection\cite{Breslau2002,Stratton2002} if finite resistivity
is introduced, changing the topology.  There can also be other
constraints that limit the accessible class of alternate
equilibria\cite{Chu2002}.  One might be able to understand the structure
of some of these possible solutions in the vicinity of the null surface
as a boundary layer analysis of a shock-like solution.  But a realistic
evaluation of such scenarios would require including finite cross-field
transport, viscosity, resistivity, and FLR effects.
We leave these issues to 
future work. Other interesting questions to consider are whether such
``non-normal'' MHD equilibria are stable to ideal and/or resistive
MHD modes and/or are experimentally accessible.
%
% GWH: sometimes have to force spacing before acknowledgements?:
% \\

%
\begin{acknowledgments}
We thank Drs.~Joshua Breslau, Ming-Sheng Chu, Nathaniel Fisch, Nikolai
Gorelenkov, Don Monticello, Paul Parks, Emilia Solano, and Leonid
Zakharov for helpful discussions on these topics.
GWH also thanks Dr. Paul Rutherford for teaching a course
that covered a virial theorem approach to the Shafranov shift.
Supported by  DOE Contract \# DE-AC02-76CH03073.

\add{{\bf NOTE:} After submitting this paper for publication, we became
aware of recent related work by Martynov {\it et
al}.\cite{Martynov03}, who study island equilibria with negative central
current in much more detail.}
\end{acknowledgments}

\bibliography{virial}% Produces the bibliography via BibTeX.

\end{document}